\documentclass[aps,prl,twocolumn,notitlepage]{revtex4-2}
\usepackage{amssymb, amsmath, amsfonts,color}
\usepackage{graphicx,xcolor}
\graphicspath{{./pics/}}
\usepackage{hyperref}

\usepackage{bbold}
\usepackage{lipsum}
\usepackage{bm}
\usepackage{natbib}
\usepackage{placeins}
\setcitestyle{square}
\usepackage{braket}

\begin{document}
  \title{Topological Nodal Point Superconductivity in Checkerboard Magnet-Superconductor Hybrid Systems}
\author{Tuan Kieu$^{1}$, Eric Mascot$^{2,3}$, Jasmin Bedow$^{1}$, Roland Wiesendanger$^{2}$ and Dirk K. Morr$^{1}$}
\affiliation{$^{1}$Department of Physics, University of Illinois at Chicago, Chicago, IL 60607, USA}
\affiliation{$^{2}$ Department of Physics, University of Hamburg, D-20355 Hamburg, Germany}
\affiliation{$^{3}$School of Physics, University of Melbourne, Parkville, VIC 3010, Australia}

\begin{abstract}
We demonstrate that checkerboard magnet-superconductor hybrid systems possess a rich phase diagram exhibiting both strong topological superconducting (STSC) and topological nodal point superconducting (TNPSC) phases. We show that TNPSC phases exist both for ferromagnetic and antiferromagnetic systems, yielding a plethora of qualitatively different edge mode structures.
Checkerboard MSH systems also facilitate the emergence of STSC phases which can be induced even in the limit of vanishing magnetization. Our results provide a new path for the quantum engineering of topological superconducting phases using atomic manipulation techniques.

\end{abstract}

\maketitle

{\it Introduction}
Magnet-superconductor hybrid (MSH) systems provide a versatile platform for the quantum engineering of topological superconductivity and the ensuing Majorana zero modes. The versatility of MSH systems arises from the fact that their topological phase diagram can be manipulated by changing the MSH system's magnetic structure. While experimentally, only two-dimensional (2D) ferromagnetic (FM) \cite{Menard2017a,Palacio-Morales2019,Kezilebieke2020} and antiferromagnetic (AFM) \cite{Bazarnik2022} MSH systems have been realized so far, intriguing topological phase diagrams -- involving strong, weak and higher-order topological superconducting phases  -- have been theoretically proposed to exist in bicollinear AFM \cite{Zhang2019a}, skyrmionic \cite{Mascot2021}, $3{\bf Q}$ \cite{Bedow2020}, and stacked magnetic structures \cite{Wong2022a}. Advances in atomic manipulation techniques \cite{Gomes2012,Polini2013} have raised the intriguing possibility to quantum engineer MSH systems \cite{Kim2018a,Palacio-Morales2019} that interpolate between these various magnetic structures, potentially giving rise to intriguing new topological phase diagrams.

\begin{figure}[htb]
  \centering
  \includegraphics[width=8cm]{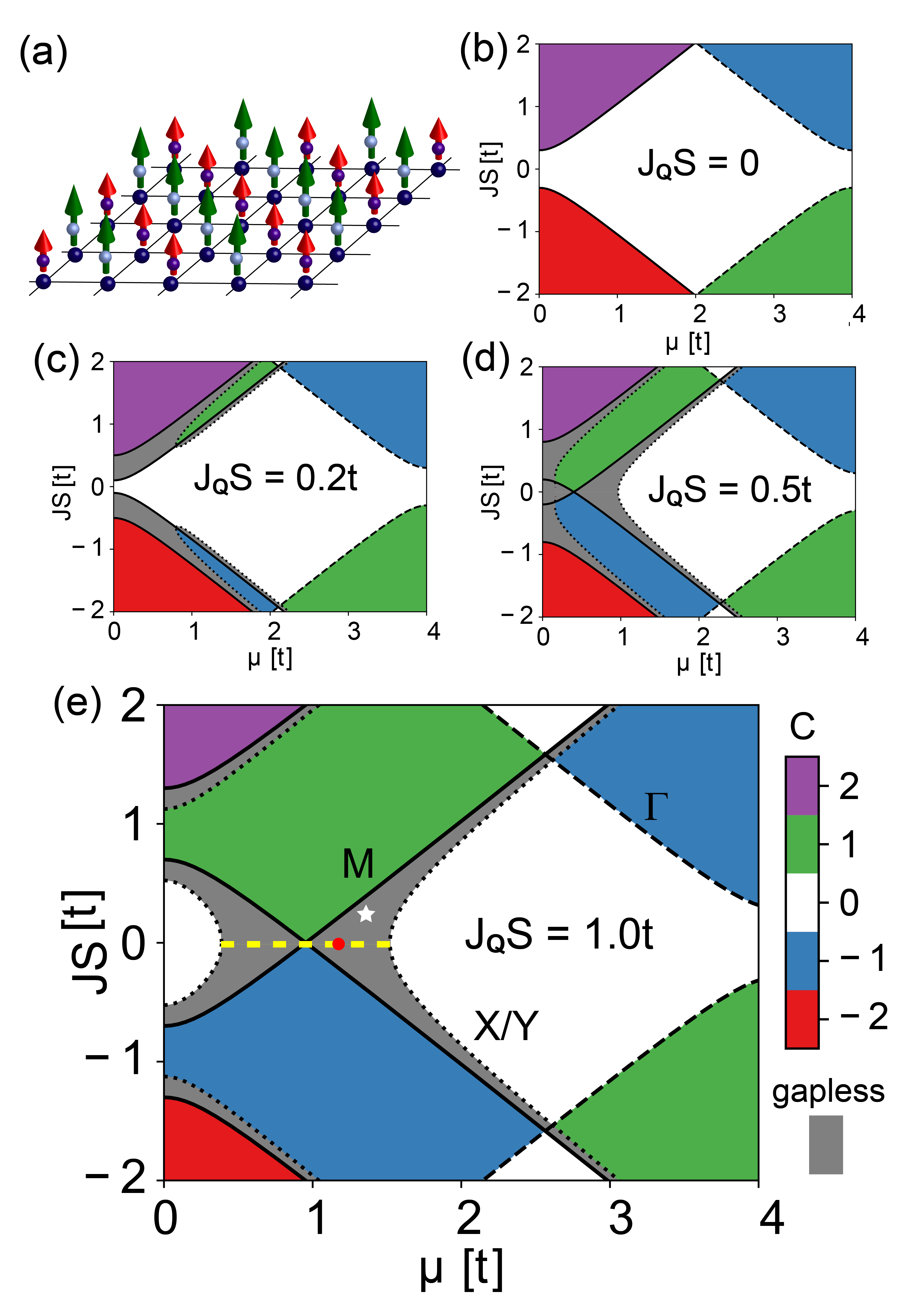}
  \caption{(a) Schematic structure of the checkerboard magnetic structure with exchange couplings $J \pm \Delta J$ in the two sublattices. (b)-(e) Topological phase diagrams of the MSH system in the $(\mu,JS)$-plane for different values of $J_{\bf Q} S$ with  $(\alpha,\Delta)=(0.2, 0.3)t$.
}
  \label{fig:Fig1}
\end{figure}

In this Letter, we consider one of these possibilities by studying MSH systems with two sublattices of different magnetic adatoms, yielding a checkerboard magnetic structure [see Fig.~\ref{fig:Fig1}(a)]. We demonstrate that such systems possess a rich topological phase diagram, exhibiting not only strong topological superconducting (STSC) phases \cite{Li2016a,Rachel2017} characterized by non-zero Chern numbers, but also extended regions of topological nodal point superconductivity (TNPSC) \cite{Baum2015a,Kao2015,Chiu2016,He2018,Zhang2019a}. The latter were recently reported to exist in the AFM MSH system  Mn/Nb(110) \cite{Bazarnik2022}, as well as in 4Hb-TaS$_2$ \cite{Nayak2021}. We show that TNPSC phases do not only exist in AFM checkerboard systems, but also in FM MSH systems where the AFM chiral symmetry is broken. Moreover, the interplay between the magnetic structures of edges and their real space direction yields a plethora of qualitatively different edge mode structures. We demonstrate that in checkerboard MSH systems STSC phases can be induced even in the limit of vanishing magnetization, in contrast to uniform FM MSH systems \cite{Li2016a,Rachel2017}. Finally, we show that TNPSC phases can be created using a single species of magnetic adatoms, simplifying their experimental realization. Our results provide a new path to quantum engineer topological superconducting phases using atomic manipulation techniques.

{\it Theoretical Model}
Starting point for our study of MSH systems with a magnetic checkerboard structure [see Fig.~\ref{fig:Fig1}(a)] is the Hamiltonian \cite{Li2016a,Rachel2017}
\begin{align}
\mathcal{H} =& \; -t \sum_{\substack{\langle{\bf r}, {\bf r}'\rangle, \\  \alpha, \beta}} c^\dagger_{{\bf r}, \alpha} c_{{\bf r}', \beta} - \mu \sum_{{\bf r}, \alpha} c^\dagger_{{\bf r}, \alpha} c_{{\bf r}, \alpha} \nonumber \\
     &+ i \alpha \sum_{{\bf r}, {\bm \delta }, \alpha, \beta} c^\dagger_{{\bf r}, \alpha}  \left({\bm \delta} \times \boldsymbol{\sigma} \right)^z_{\alpha, \beta}   c_{{\bf r + \delta}, \beta} \nonumber \\
    & - \Delta \sum_{{\bf r}} \left( c^\dagger_{{\bf r}, \uparrow} c^\dagger_{{\bf r}, \downarrow} + c_{{\bf r}, \downarrow} c_{{\bf r}, \uparrow} \right) \nonumber \\
    &-  {\sum_{{\bf r} , \alpha, \beta}}  \left(J + J_{\bf Q} e^{i {\bf Q}\cdot {\bf r}} \right) c^\dagger_{{\bf r}, \alpha}  {\bf S}_{\bf r} \cdot \boldsymbol{\sigma}_{\alpha,\beta} c_{{\bf r}, \beta} \; .
    \label{eq:H}
\end{align}
Here, the operator $c^\dagger_{{\bf r}, \alpha}$ creates an electron of spin $\alpha$ at site ${\bf r}$, $-t$ is the nearest-neighbor hopping amplitude on a two-dimensional square lattice, $\mu$ is the chemical potential, $\alpha$ is the Rashba spin-orbit coupling between nearest-neighbor sites ${\bf r}$ and ${\bf r} +{\bm \delta}$, and $\Delta$ is the $s$-wave superconducting order parameter. The last term in Eq.(\ref{eq:H}) describes the coupling between the adatoms' spin ${\bf S}_{\bf r}$  of magnitude $S$  at site ${\bf r}$ and the conduction electrons, with exchange couplings $J \pm J_{\bf Q}$ in the two sublattices, and ${\bf Q}=(\pi,\pi)$ [for details, see Supplemental Material (SM) Sec.I].  Due to the hard superconducting gap, which suppresses Kondo screening, we can consider the spins of the magnetic adatoms to be classical in nature \cite{Balatsky2006,Heinrich2018}.

To characterize the strong topological phases, we compute its topological invariant, the Chern number, via \cite{Avron1983}
\begin{align}\label{eq:C}
 C & =  \frac{1}{2\pi i} \int_{\text{BZ}} d^2k \mathrm{Tr} ( P_{\bf{k}} [ \partial_{k_x} P_{\bf{k}}, \partial_{k_y} P_{\bf{k}} ] )  \nonumber \\
 P_{\bf{k}} & = \sum_{E_n(\bf{k}) < 0} |\Psi_n({\bf{k}}) \rangle \langle \Psi_n({\bf{k}})|
\end{align}
where $E_n({\bf{k}})$ and $|\Psi_n({\bf{k}}) \rangle$ are the eigenenergies and the eigenvectors of the Hamiltonian in Eq.(\ref{eq:H}), with $n$ being a band index, and the trace is taken over Nambu, spin and sublattice space. Since STSC phases occur only for $J \not = 0$, they belong to the topological class $D$ \cite{Kitaev2009,Ryu2010,Chiu2016}.

Finally, for antiferromagnetic TNPSC phases ($J = 0$) the chiral symmetry of the system allows us to compute a topological charge and characteristic angle associated with nodal points (see SM Sec.~II). We note that trivial nodal point superconductivity was previously discussed in the context of antiferromagnetic semimetals \cite{Brzezicki2018}.

\begin{figure}[t]
  \centering
  \includegraphics[width=8.5cm]{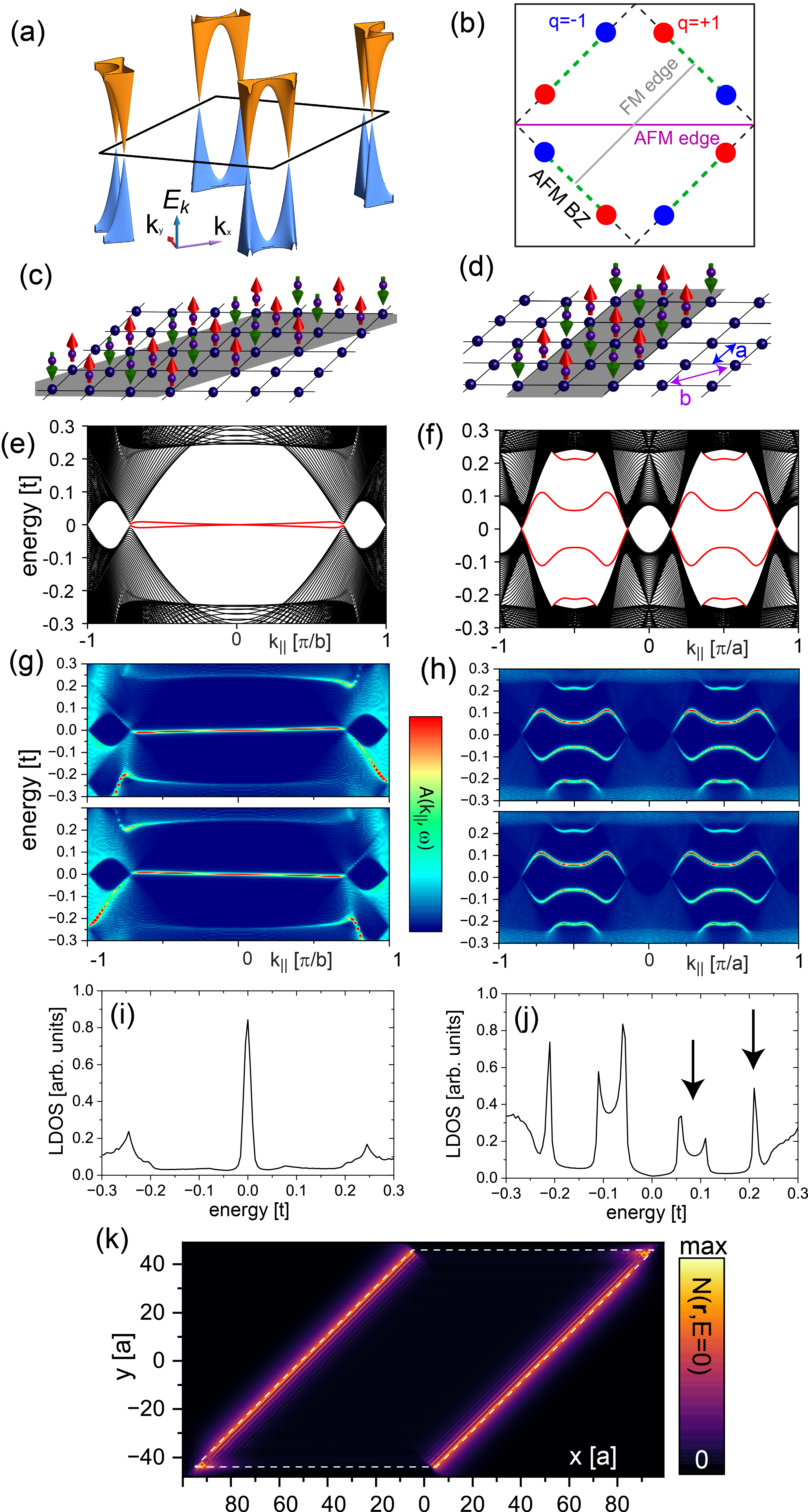}
  \caption{ (a) Energy dispersion of the bulk system in the magnetic BZ. (b) Nodal points and edge modes (dashed green lines) projected onto momenta parallel to the FM and AFM edges. Schematic representation of (c) diagonal (FM) and (d) vertical (AFM) edges. Electronic structure of a ribbon with (e) FM and (f) and AFM edges. (g),(h) Spectral functions at the ribbon's left (upper panel) and right (lower panel) edges corresponding to (e) and (f). (i), (j) LDOS at the edges corresponding to (e) and (f). (k) Zero-energy LDOS of a magnetic island (dashed white line) with FM and AFM edges.  $(\mu, \alpha, \Delta, JS, J_{\bf Q}S)=(1.2, 0.2, 0.3, 0, 1.0)t.$
}
  \label{fig:Fig2}
\end{figure}
{\it Results}
We begin by considering the topological phase diagram of the Hamiltonian in Eq.(\ref{eq:H}), which is presented in Figs.~\ref{fig:Fig1}(b)-(e) for different values of $J_{\bf Q}S$ (the  qualitative form of these phase diagrams is independent of the particular values of $\alpha$ and $\Delta$, see SM Sec.~III). In addition to the strong topological $C=2,-1$ phases which are already present for $J_{\bf Q}S = 0$, two new features appear in the phase diagram with increasing $J_{\bf Q}S$. First, a $C=1$ phase emerges due to the backfolding of the Brillouin zone (BZ), and the splitting of the bands at the $X/Y$-points. Secondly, an extended region of the phase diagram (shown in grey), lying between the gap closing lines of the $M$- (solid black line) and $X/Y$-points (dotted black line), becomes gapless (an analytic expression for the phase transition lines is given in SM Sec.~IV). As one moves in the gapless region between these two gap closing lines, the nodal points move along the magnetic BZ boundary from the $M$ to the $X/Y$-points. For $J_{\bf Q}S \geq \sqrt{2 \alpha^2 + \Delta^2_0}$, the strong topological $C=\pm1$ phases touch at a single antiferromagnetic ($JS=0$) point in the phase diagram with chemical potential $\mu_{c} = \sqrt{(J_{\bf Q}S)^2-\Delta^2}$. At this point, any non-zero magnetization, which can be induced by arbitrarily small magnetic fields, pushes the system into a STSC  phase, in contrast to uniform FM MSH systems, where $JS > \Delta$ is required \cite{Li2016a,Rachel2017} for the emergence of strong topological phases.

To exemplify the electronic structure of an antiferromagnetic MSH system ($JS=0$) in the gap-less region around $\mu_c$ [yellow line in Fig.~\ref{fig:Fig1}(e)], we consider a system with $\mu=1.2t$ [red dot in Fig.~\ref{fig:Fig1}(e)].  In Fig.~\ref{fig:Fig2}(a) we present the resulting electronic structure exhibiting eight nodal points located along the boundary of the magnetic BZ. Due to the chiral symmetry of the system along the $JS=0$ line, we find that these nodal points possess a non-zero topological charge $q=\pm 1$ [see Fig.~\ref{fig:Fig2}(b)], the corresponding characteristic angles are shown in SM Sec.~II.
The presence of nodal points with opposite topological charges allows for the emergence of edge modes connecting such nodal points [schematically shown as dashed green lines in Fig.~\ref{fig:Fig2}(b)] in MSH systems exhibiting edges \cite{Baum2015a,Kao2015,Chiu2016,He2018,Zhang2019a}. When considering MSH systems in a ribbon geometry possessing diagonal ferromagnetic (FM) or vertical/horizontal antiferromagnetic (AFM) edges [see Figs.~\ref{fig:Fig2}(c) and (d)], we find that their electronic structures indeed exhibit such in-gap edge modes as shown by red lines in Figs.~\ref{fig:Fig2}(e) and (f), respectively. However, in contrast to an MSH system with AFM edges, the edge modes along FM edges disperse only weakly around zero-energy. To understand this qualitative difference, we note that the projection of the bulk band structure onto momenta parallel to either an FM or AFM edge leads to two overlapping (in momentum space) edge modes (see Fig.~\ref{fig:Fig2}(b), and SM Sec.~V). These two edge modes can in general hybridize, and thus split in energy. The spectral functions at the FM and AFM edges, shown in the upper and lower panels of Figs.~\ref{fig:Fig2}(g) and (h), respectively, for the two edges of each ribbon, provide insight into the different nature of the resulting edge modes. In contrast to the AFM edge, the edge mode along the FM edge is chiral in nature \cite{Rachel2017} (due to the broken time-reversal symmetry along the edge) with opposite Fermi velocities on the two FM edges, as shown in Fig.~\ref{fig:Fig2}(g). Thus, the two modes along the FM edges are spatially separated, and thus do not hybridize. In contrast, the spectral functions at the AFM edges [see Fig.~\ref{fig:Fig2}(h)] show that both modes exist at the same edge  and thus strongly hybridize, leading to the much larger energy splitting. This qualitative difference leads to distinct signatures in the local density of states (LDOS) of FM and AFM edges: while the LDOS at an FM edge exhibits a large zero-energy peak [see Fig.~\ref{fig:Fig2}(i)], the LDOS at the AFM edge exhibits only a low-energy V-like shaped LDOS [see Fig.~\ref{fig:Fig2}(j)], with the signatures of the edge modes seen only at higher energies [see black arrows]. As a result, a spatial plot of the zero-energy LDOS of a finite magnetic island [see Fig.~\ref{fig:Fig2}(k)] reveals large spectral weight along the FM edges, and vanishing spectral weight along the AFM edges.

\begin{figure}[htb]
  \centering
  \includegraphics[width=8.5cm]{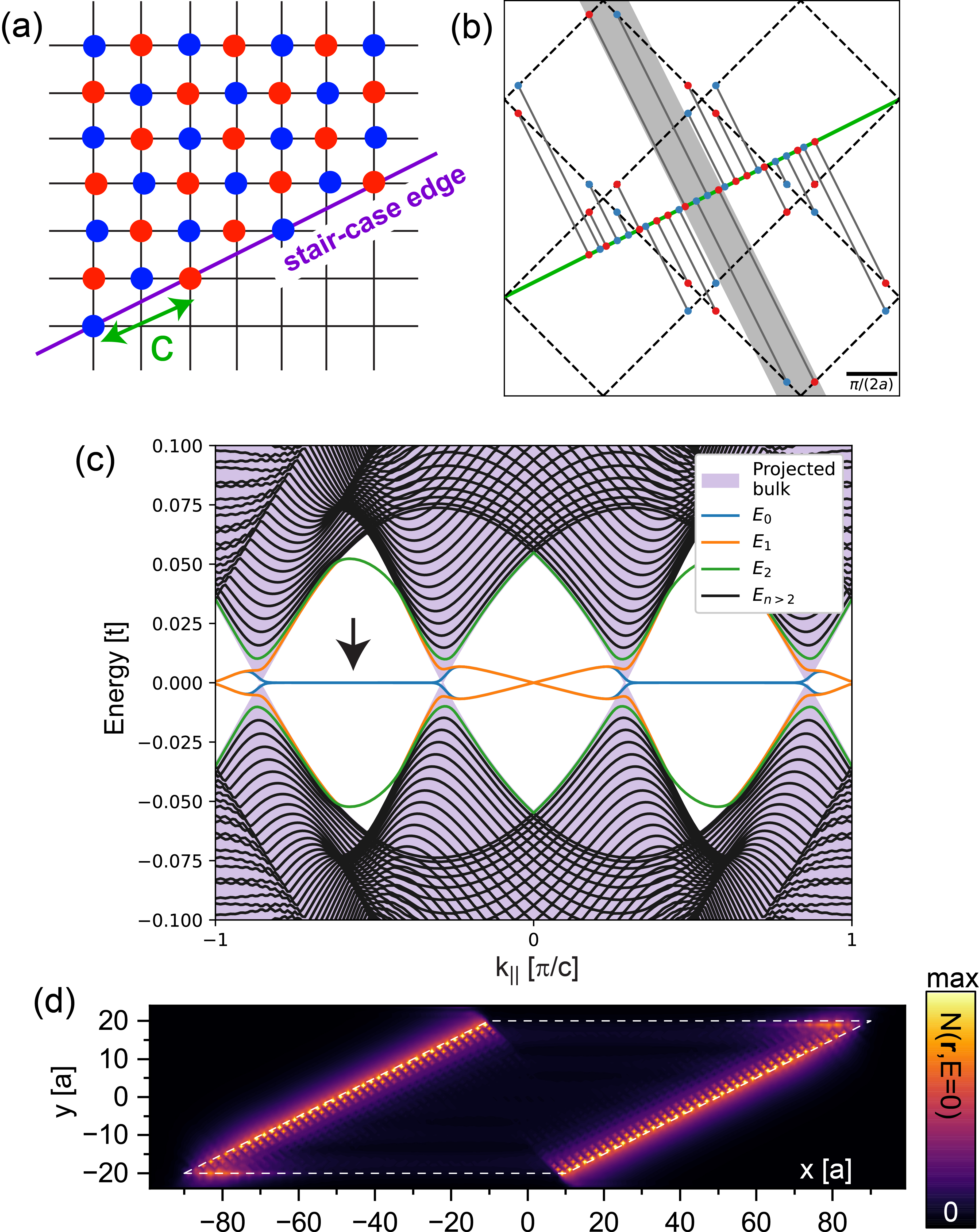}
  \caption{(a) Schematic picture of a stair-case edge. (b) Magnetic BZ and nodal points projected onto the momentum parallel to the stair-case edge. (c)  Electronic structure of the ribbon. (d) Zero-energy LDOS for a magnetic island with horizontal and stair-case edges. Parameters are $(\mu, \alpha, \Delta, JS, J_{\bf Q}S)=(1.2, 0.2, 0.3, 0, 1.0)t.$}
  \label{fig:Fig5}
\end{figure}
To obtain flat, non-dispersive edge modes, it is necessary to consider edges in real space such that the projected edge modes do not completely overlap in momentum space, as is the case for the FM and AFM edges (see SM Sec.~V). For example, by considering a stair-case like edge in real space [see Fig.~\ref{fig:Fig5}(a)], we find that the resulting projection of the nodal points onto momenta parallel to the stair case edge [Fig.~\ref{fig:Fig5}(b)] yields momentum regions with even and odd numbers of projected modes. In those regions where three projected modes overlap, one edge mode is non-dispersive and located at zero energy [see black arrows in Fig.~\ref{fig:Fig5}(c)], while the other two projected modes hybridize and thus shift to higher energies. As a result, a magnetic island that possesses both vertical and stair-case edges [see Fig.~\ref{fig:Fig5}(d)] again shows a large zero-energy LDOS along the stair-case edges, and vanishing spectral weight along the vertical edges.

\begin{figure}[htb]
  \centering
  \includegraphics[width=8cm]{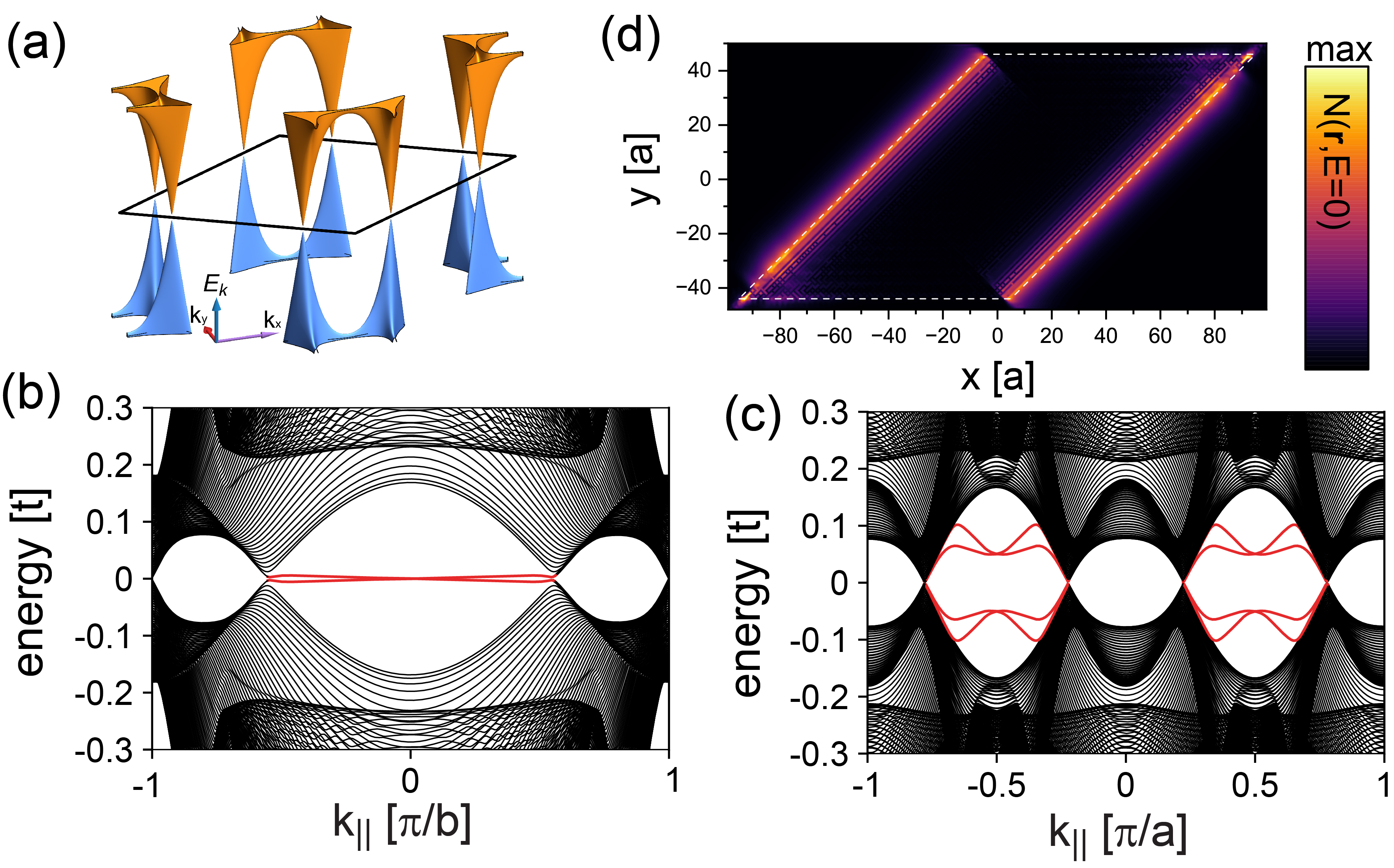}
  \caption{(a) Energy dispersion of the bulk system in the magnetic BZ.  Electronic structure of a ribbon with (b) FM and (c) AFM edges, as a function of momentum parallel to the edge. (k) Zero-energy LDOS of a magnetic island (dashed white line) with FM and AFM edges.  $(\mu, \alpha, \Delta, JS,J_{\bf Q}S)=(1.4, 0.2, 0.3, 0.25, 1.0)t.$
}
  \label{fig:Fig3}
\end{figure}
The question immediately arises whether the nature of the edge modes persists in those gapless regions of the phase diagram where the chiral symmetry of the AFM is broken due to a non-zero $JS$. To answer this question, we consider a system in the gapless region above the  antiferromagnetic line, as indicated by a white star in Fig.~\ref{fig:Fig1}(e). The electronic bulk structure of this system again reveals 8 nodal points along the magnetic BZ boundary, as shown in Fig.~\ref{fig:Fig3}(a). The electronic dispersions for ribbons with FM and AFM edges [Figs.~\ref{fig:Fig3}(b) and (c), respectively] reveal the same qualitative nature of the respective edge modes as in the $JS=0$ case discussed in Fig.~\ref{fig:Fig2}. However, since the chiral symmetry of the AFM edge is broken due to the non-zero $JS$, the double-degeneracy of the AFM edge modes is lifted, leading to a further energy splitting among them. The zero-energy LDOS of a magnetic island [see Fig.~\ref{fig:Fig3}(d)] thus again reveals large spectral weight along the FM edges, and vanishing spectral weight along the AFM edges. Similar results are obtained not only for any point in the gapless regions around $\mu_c$ [see Fig.~\ref{fig:Fig1}(e)], but also in the gapless regions induced by much smaller values of $J_{\bf Q}S$ [see Figs.~\ref{fig:Fig1}(c),(d)], we conclude that checkerboard MSH systems are ideally suited to quantum engineer topological nodal point superconductivity as the occurrence of TNPSC phases does not require the fine-tuning of parameters. Of particular interest here are MSH systems with large Rashba spin-orbit coupling since the extent of gapless regions in the phase diagram increases with $\alpha$ (see SM Sec.~III). Our findings provide further support for the conclusion of recent scanning tunneling spectroscopy (STS) experiments \cite{Bazarnik2022}  that the presence or absence of (near) zero-energy edge modes at various edges in the AFM MSH system  Mn/Nb(110) are characteristics signatures of the underlying TNPSC phase.

\begin{figure}[htb]
  \centering
  \includegraphics[width=8cm]{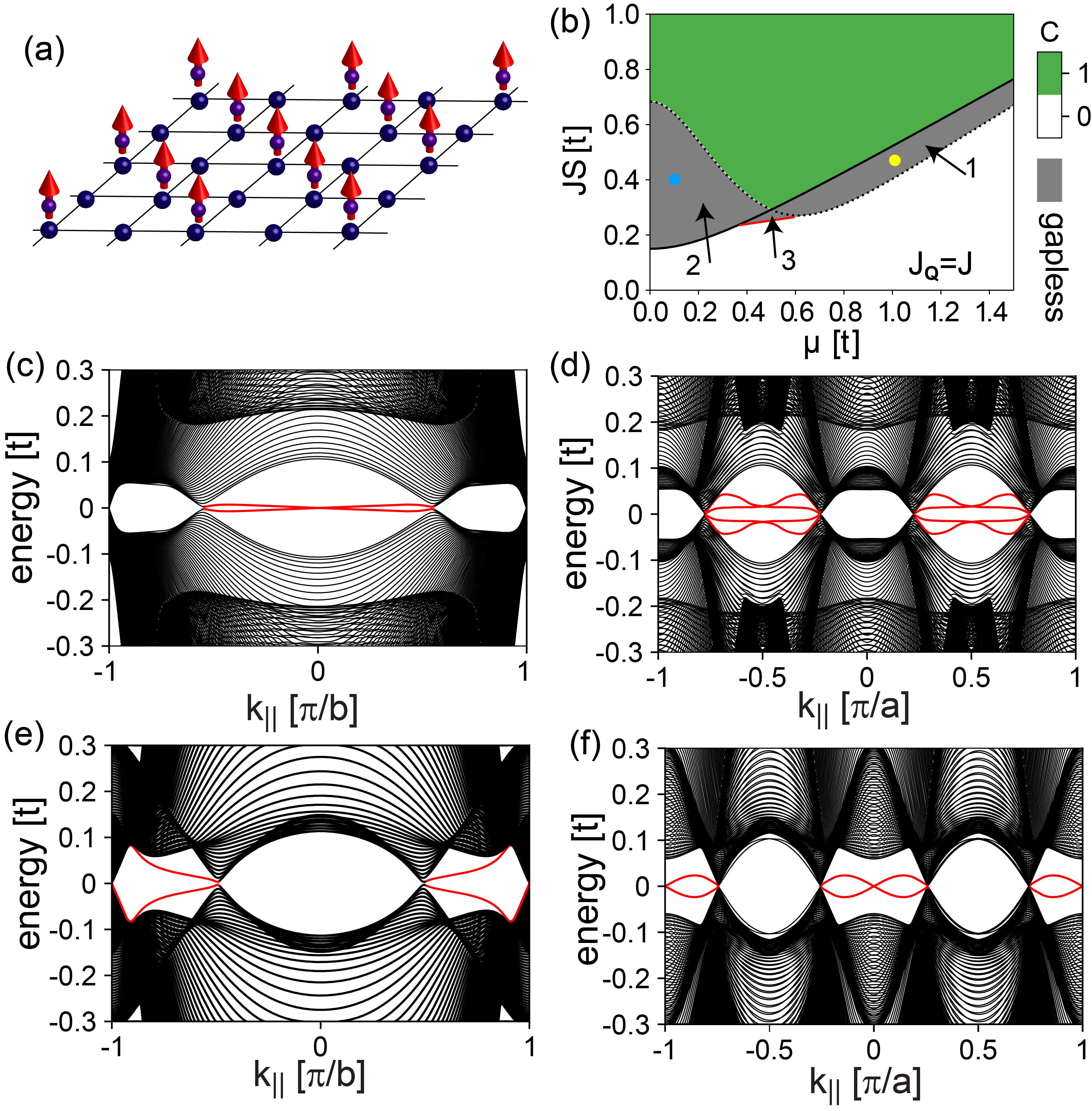}
  \caption{(a) Schematic structure of the checkerboard magnetic structure for $J = J_{\bf Q}$. (b) Topological phase diagram for $(\alpha, \Delta)=(0.2, 0.3)t$. Electronic structure of an MSH ribbon with (c),(e) FM and (d)(f) AFM edges for $(\mu,JS)=(1.0,0.47)t$ [yellow dot in (b)], and $(\mu,JS)=(0.1,0.4)t$ [blue dot in (b)], respectively.
}
  \label{fig:Fig4}
\end{figure}

A special case of the magnetic checkerboard structure arises when $J=J_{\bf Q}$, implying that only a single species of magnetic adatoms is present in one of the two sublattices, as schematically shown in Fig.~\ref{fig:Fig4}(a).
In Fig.~\ref{fig:Fig4}(b) we present the resulting phase diagram. In addition the strong topological $C=1$ phase, we obtain three types of gapless regions. Region 1 is entered from the gapped trivial ($C=0$) region via a gap closing at the $X/Y$-points. As a result, the edge modes connect the nodal points of opposite topological charge that are located on the same edge of the magnetic BZ. In this region, the electronic structure along FM and AFM edges [see Figs.~\ref{fig:Fig4}(c) and (d)] is similar to that shown in Fig.~\ref{fig:Fig2}. In contrast, the transition into region 2 from the trivial phase requires a gap closing at the $M$-points, which yields edge modes connecting the nodal points of opposite topological charge on neighboring edges of the magnetic BZ [see Figs.~\ref{fig:Fig4}(e) and (f)].  In this case, neither edge shows a (nearly) non-dispersive edge mode, implying the absence of any pronounced zero-energy peak in the LDOS at the edges. Finally, the transition line into region 3 (red line) shows gap closings that vary along the magnetic BZ boundary between the $M$ and $X/Y$-points. We thus conclude that by even using a single species of magnetic adatoms, it is possible to create a TNPSC phase.

{\it Discussion} We have shown that checkerboard MSH systems are ideally suited to quantum engineer STSC as well as TNPSC phases, with the latter existing both for AFM ($J=0$) and FM ($J \not =0$) structures. The nature of edge modes in the TNPSC phases is sensitively determined by the projection of nodal points onto the edge momenta, and the magnetic structure of the edge. This allows for the emergence of edge modes that are dispersive in some momentum ranges, and non-dispersive (flat) in others. Our results further support the conclusions of recent STS experiments in the AFM MSH system  Mn/Nb(110) \cite{Bazarnik2022} that the underlying TNPSC phase can be identified through the presence or absence of (near) zero-energy edge modes along different edges in the system.
The finding that TNPSC phases can be created even with a single species of magnetic adatoms when placed in only one of the two sublattices further supports the ubiquity of TNPSC phases in checkerboard MSH systems. Advances in atomic manipulation techniques all but ensure that such checkerboard MSH systems and the ensuing TNPSC phases can be quantum engineered in the near future.

\section{Acknowledgments}
The authors would like to thank S. Rachel for stimulating discussions. T.K, E.M, J.B., and D.K.M. acknowledge support by the U. S. Department of Energy, Office of Science, Basic Energy Sciences, under Award No. DE-FG02-05ER46225.
R.W. acknowledges financial support by the EU via the ERC Advanced Grant ADMIRE (No. 786020) and the DFG via the Cluster of Excellence “Advanced Imaging of Matter” (EXC 2056, Project ID 390715994).

\end{document}


\preprint{APS/123-QED}

  \title{Topological Nodal Point Superconductivity in Checkerboard Magnet-Superconductor Hybrid Systems \\[0.25cm]
{\large Supplementary Material}}
\author{Tuan Kieu$^{1}$, Eric Mascot$^{2,3}$, Jasmin Bedow$^{1}$, Roland Wiesendanger$^{2}$ and Dirk K. Morr$^{1}$}
\affiliation{$^{1}$Department of Physics, University of Illinois at Chicago, Chicago, IL 60607, USA}
\affiliation{$^{2}$Department of Physics, University of Hamburg, D-20355 Hamburg, Germany}
\affiliation{$^{3}$School of Physics, University of Melbourne, Parkville, VIC 3010, Australia}

\maketitle

\section{Hamiltonian in momentum space}

For a checkerboard MSH system, the Hamiltonian of Eq.(1) in the main text can be written in momentum space as $ {\cal H} = \Psi^{\dagger}_{\bf k} {\hat H}_{\bf k} \Psi_{\bf k}$ with spinor
\begin{align*}
  \Psi^{\dagger} = \left[ c^{\dagger}_{k,\uparrow},
        c^{\dagger}_{k+Q,\uparrow},
        c^{\dagger}_{k,\downarrow},
        c^{\dagger}_{k+Q,\downarrow},
        c_{-k,\downarrow},
        c_{-k+Q,\downarrow},
        -c_{-k,\uparrow},
        -c_{-k+Q,\uparrow}
\right]
\end{align*}
and the Hamiltonian matrix ${\hat H}_{\bf k}$ is given by
\begin{align}
        \label{eq:H-matrix}
        {\hat H}_{\bf k} = \begin{pmatrix}
                \epsilon_k-JS & -J_{\bf Q}S & -A_k & 0 & -\Delta & 0 & 0 & 0 \\
                -J_{\bf Q}S & \epsilon_{k+Q}-JS & 0 & -A_{k+Q} & 0 & -\Delta & 0 & 0 \\
                -A^*_k & 0 & \epsilon_k+JS & J_{\bf Q}S &0  & 0 & -\Delta & 0 \\
                0 & -A^*_{k+Q} & J_{\bf Q}S & \epsilon_{k+Q}+JS & 0 & 0 & 0 & -\Delta \\
                -\Delta & 0 & 0 & 0 & -\epsilon_{-k}-JS & -J_{\bf Q}S & -A_{-k} & 0 \\
                0 & -\Delta & 0 & 0 & -J_{\bf Q}S & -\epsilon_{-k+Q}-JS & 0 & -A_{-k+Q} \\
                0 & 0 & -\Delta & 0 & -A^*_{-k} & 0 & -\epsilon_{-k}+JS & J_{\bf Q}S \\
                0 & 0 & 0 & -\Delta & 0 & -A^*_{-k+Q} & J_{\bf Q}S & -\epsilon_{-k+Q}+JS
        \end{pmatrix} \ ,
\end{align}
where $\epsilon_k=-2t(\cos k_x +\cos k_y)-\mu$ and $A_k = -2\alpha(\sin k_x +i\sin k_y)$ is the Rashba spin-orbit interaction.

Note that the spatial localization length of the (near) zero-energy edge modes is determined by the superconducting coherence length, $\xi_c$, which decreases with increasing $\Delta$. Thus in order to demonstrate the localization of edge modes along the FM edges in finite size magnetic islands [see, e.g., Fig.~2(k) of the main text], $\xi_c$ has to be chosen significantly smaller than the linear dimensions of the real space systems we can consider (which is limited by computational resources). This resulted in the choice of a somewhat large superconducting gap of $\Delta=0.3t$ for the calculations shown in the main text. However, using smaller values of $\Delta$ does not change the qualitative form of the phase diagram (see SM Sec.~III), and thus does not affect the conclusions presented in the main text.

\section{Topological charge of nodal points}

For an antiferromagnetic MSH systems with $JS=0$, the Hamiltonian in Eq.~(\ref{eq:H-matrix}) possesses an effective time-reversal $T = \tau_0 \sigma_y \lambda_z K$, particle-hole $C = \tau_y \sigma_y \lambda_0 K$, and chiral symmetry $S = \tau_y \sigma_0 \lambda_z$ where $\tau_a$, $\sigma_b$, $\lambda_c$ are Pauli matrices in particle-hole, spin, and sublattice space, respectively, and $K$ is the complex conjugate operator. The symmetries square to $T^2=-1$ and $C^2=S^2=1$, yielding the topological symmetry class DIII.
To transform the Hamiltonian in Eq.~(\ref{eq:H-matrix}) to an off-diagonal block form, we choose a basis in which $S$ is diagonal, yielding
\begin{align}
    U^\dagger S U =& \begin{pmatrix} 1 & 0 \\0 & -1 \end{pmatrix} &
   {\hat H}_{\bf k}^\prime =  U^\dagger {\hat H}_{\bf k} U =& \begin{pmatrix} 0 & h_{\bf k} \\ h_{\bf k}^\dag & 0 \end{pmatrix}.
    \label{eq:H-block}
\end{align}
Next, we define,
\begin{align}
    Q_{\bf k} = \sum_n \ket{n_{\bf k}} {\rm sgn}(E_{n,\bf k}) \bra{n_{\bf k}}
    = \begin{pmatrix} 0 & q_{\bf k} \\ q_{\bf k}^\dag & 0 \end{pmatrix},
\end{align}
where $E_{n,\bf k}$ are the energies and $\ket{n_{\bf k}}$ are the eigenvectors of ${\hat H}_{\bf k}^\prime$ in Eq.~(\ref{eq:H-block}).
The characteristic angle, $\theta_{\bf k}$, is then defined via $e^{i \theta_{\bf k}} = \det(q_{\bf k})$.
The topological charge of each nodal point is the winding number of the characteristic angle around the nodal point, which is expressed as
\begin{align}
    \nu = \frac{1}{2\pi i} \oint d{\bf k} \cdot {\rm Tr} \left[
        q_{\bf k}^{-1} \nabla_{\bf k} q_{\bf k}
    \right]
    = \frac{\Delta \theta_{\bf k}}{2\pi}.
\end{align}

\begin{figure}[htb]
  \centering
  \includegraphics[width=8cm]{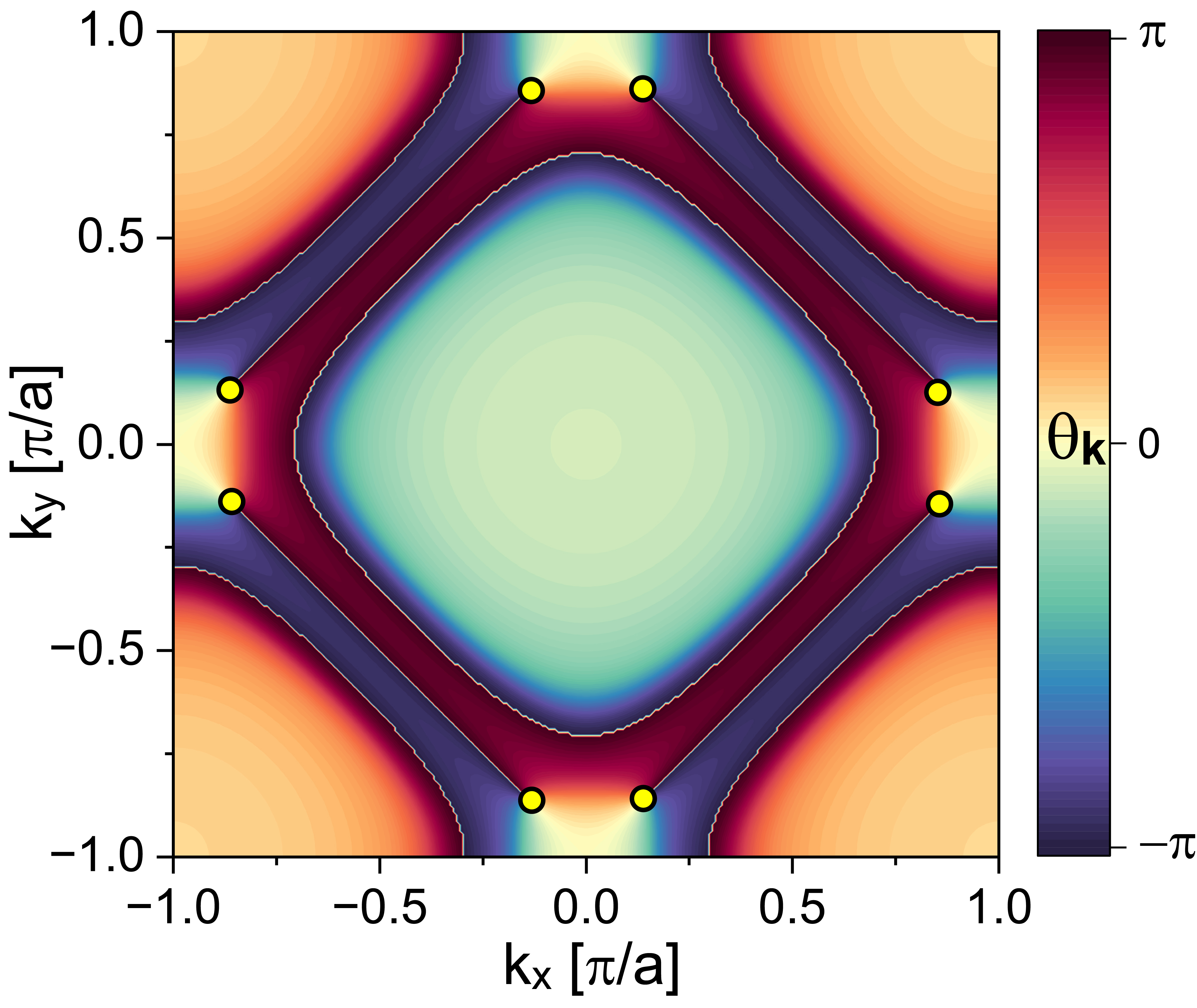}
  \caption{Characteristic angle $\theta_{\bf k}$ in the BZ for parameters $(\mu, \alpha, \Delta, JS, J_{\bf Q}S)=(1.2, 0.2, 0.3, 0, 1.0)t$. The nodal points shown as filled yellow circles with black edges are connected by a branch cut. }
  \label{fig:SIFig2}
\end{figure}

\section{Topological Phase Diagram}

\begin{figure}[htb]
  \centering
  \includegraphics[width=16cm]{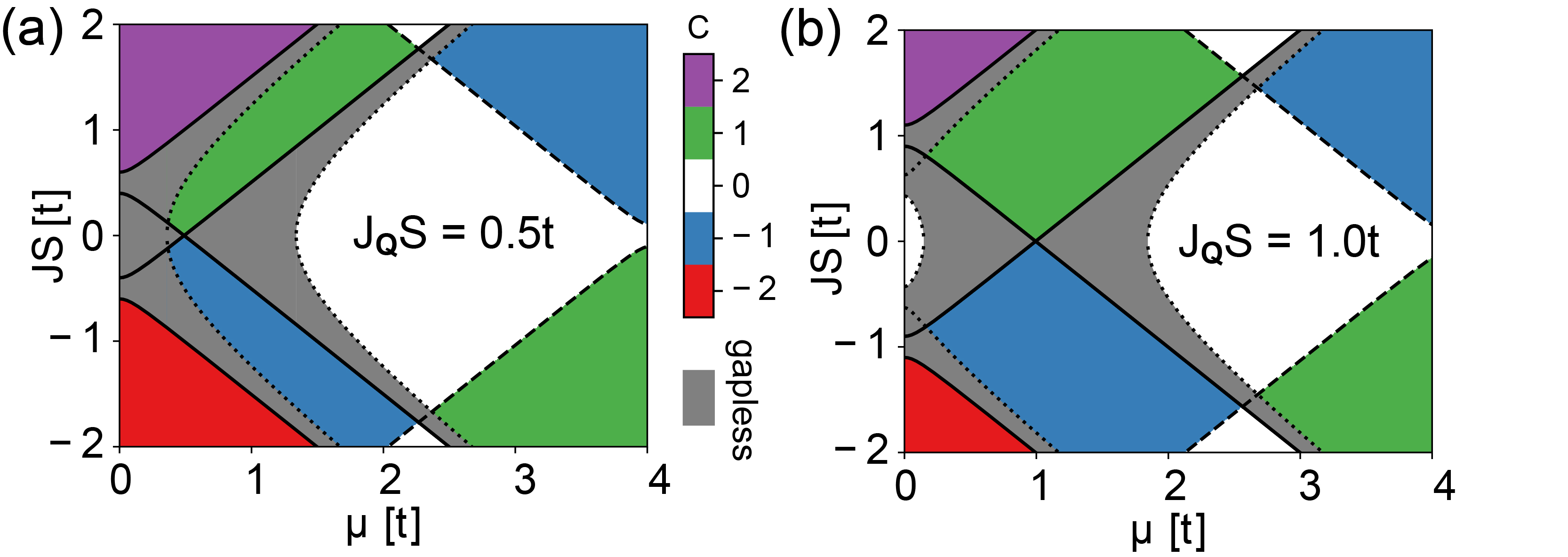}
  \caption{Topological phase diagram in the $(\mu, JS)$-plane with $(\alpha, \Delta)=(0.3,0.1)t$ and (a) $J_{\bf Q}S=0.5t$, and  (b) $J_{\bf Q}S=1.0t$. }
  \label{fig:SIFig3}
\end{figure}

The topological phase diagrams shown in Fig.~1 of the main text remain qualitatively unchanged when $\alpha$ and $\Delta$ are changed. To demonstrate this, we present in Fig.~\ref{fig:SIFig3} the topological phase diagram for $(\alpha, \Delta)=(0.3,0.1)t$ and two different values of $J_{\bf Q}S$,  which contain the same topological phases and phase transitions as the phase diagrams shown in Fig.~1 of the main text.
Note, however, that the extent of the gapless TNPSC regions in the phase diagram increases with increasing Rashba spin-orbit coupling, making superconductors with a large Rashba spin-orbit coupling particularly interesting for the realization of checkerboard MSH systems.

\section{Analytic expression for the phase transition lines}

\begin{figure}[htb]
  \centering
  \includegraphics[width=6cm]{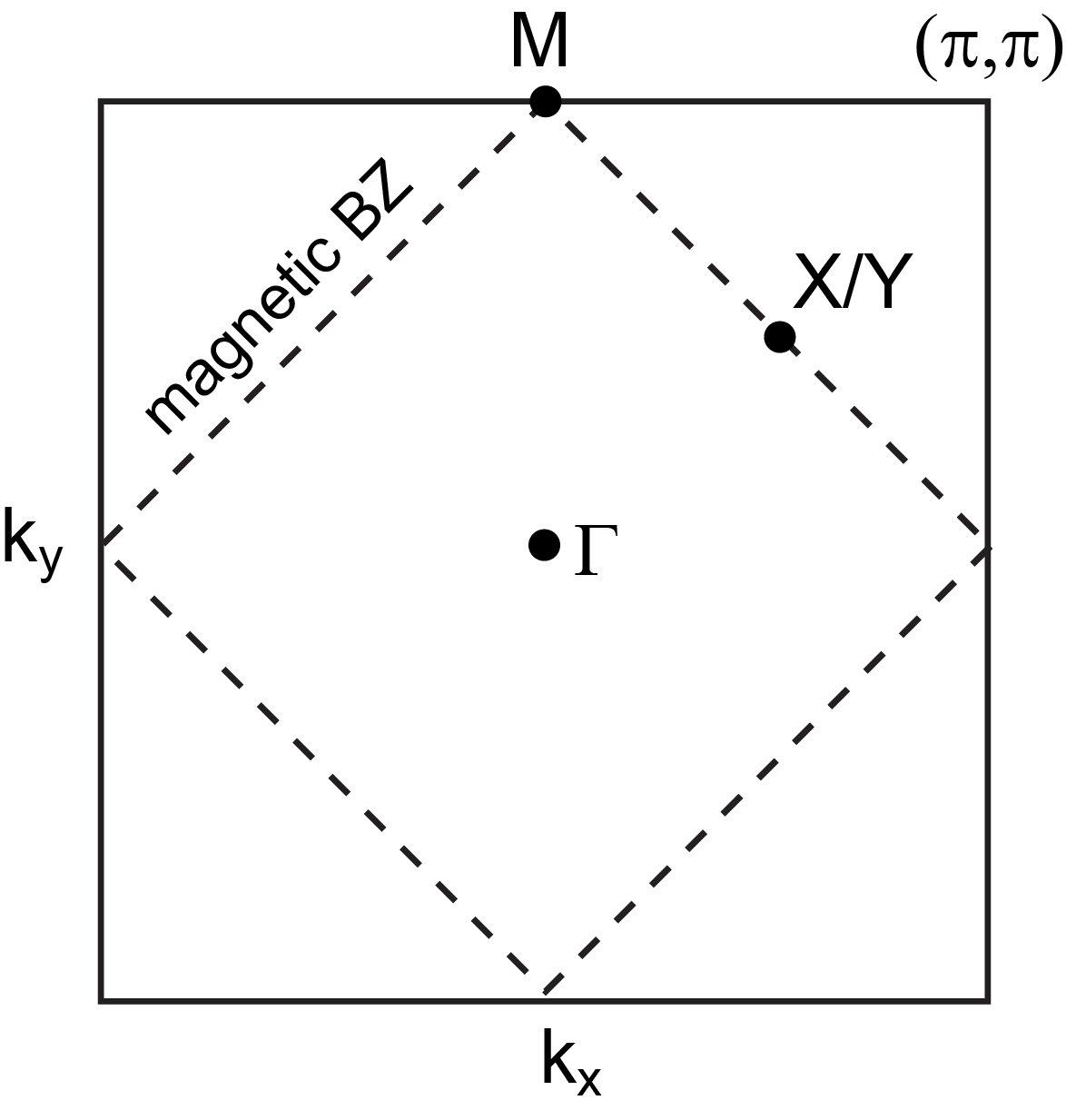}
  \caption{High symmetry points of the magnetic Brillouin zone.}
  \label{fig:SIFig1}
\end{figure}
The various phase transition lines shown in Fig.1 of the main text are accompanied by gap closings at high symmetry points in the magnetic Brillouin zone -- the $\Gamma$-, $X/Y$-, and  $M$-points (see Fig.~\ref{fig:SIFig1})-- which are described by the relations
\begin{align}
  \label{eq:PT1}
   JS &= \pm \left[ 16t^2+(J_{\bf Q}S)^2+\Delta^2+\mu^2 \pm
  2\sqrt{16t^2\mu^2+(J_{\bf Q}S)^2(\Delta^2+\mu^2)} \right]^{1/2} \nonumber \\
\end{align}
at the $\Gamma$-point (dashed black line in Fig.1 of the main text), by
\begin{align}
  \label{eq:PT2}
 JS &=\pm J_{\bf Q}S\pm\sqrt{\Delta^2+\mu^2}\nonumber\\
\end{align}
at the $M$-point (solid black line in Fig.1 of the main text), and by
\begin{align}
  \label{eq:PT3}
 JS &= \pm \left[ (J_{\bf Q}S)^2 - 8\alpha^2+\Delta^2+\mu^2  \pm
  2\sqrt{(J_{\bf Q}S)^2\Delta^2-8\alpha^2\Delta^2+(J_{\bf Q}S)^2\mu^2} \right]^{1/2} \nonumber \\
\end{align}
at the  $X/Y$-points (dotted black line in Fig.1 of the main text).

\section{Projections of nodal points and emergence of edge modes }

\begin{figure}[htb]
  \centering
  \includegraphics[width=12cm]{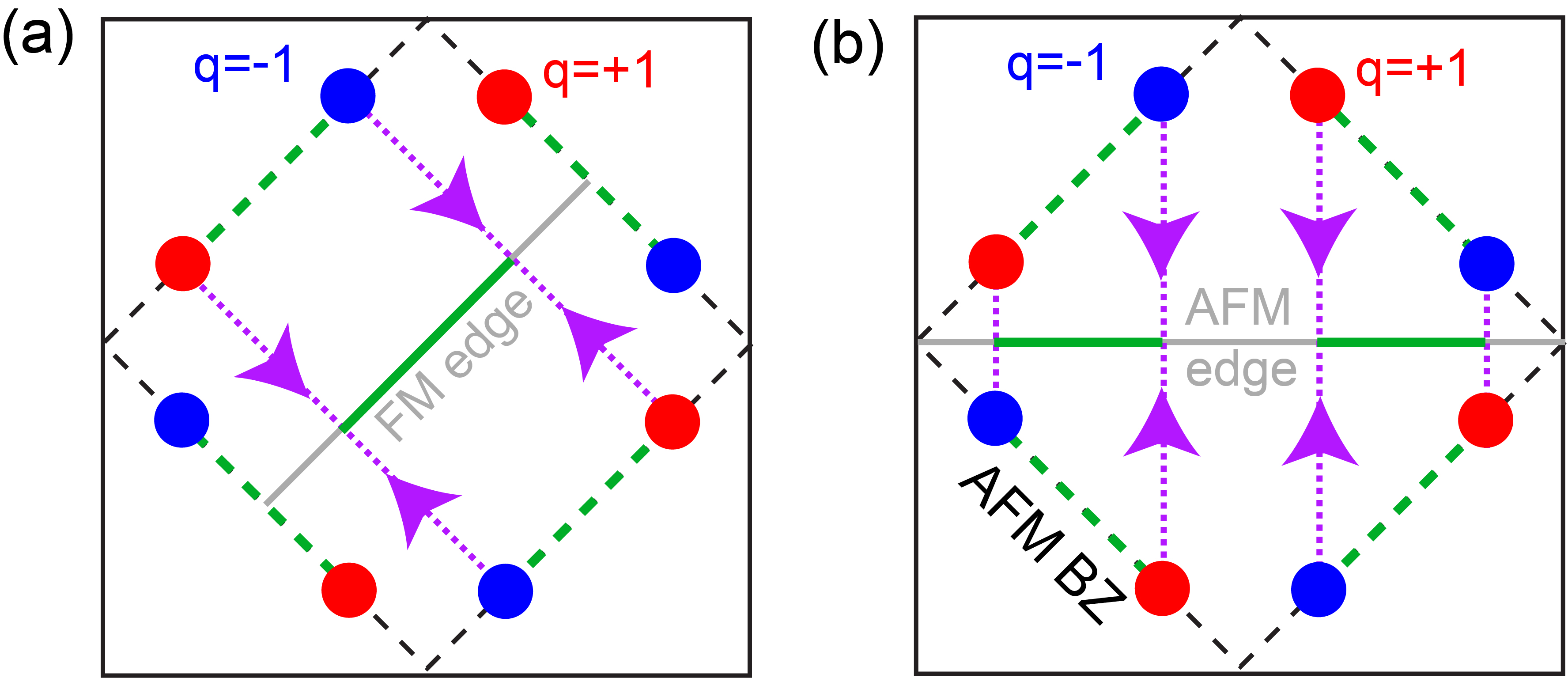}
  \caption{Nodal points and their projections onto momenta in the magnetic BZ (dashed black line) along (a) a FM edge, and (b) an AFM edge. The dashed green lines represent the edge modes connecting the nodal points of opposite topological charge, while the solid green lines represent their projections onto the edge momenta. }
  \label{fig:SIFig2}
\end{figure}
To understand the form of edge modes along FM and AFM edges in more detail, we consider the projection of the nodal points onto a momentum line parallel to the edges, as schematically shown in Fig.~\ref{fig:SIFig2}. If the gapless TNPSC phase is entered from the gapped trivial phase via a gap closing at the $X/Y$-points, the edge modes (dashed green lines) connect nodal points of opposite topological charge that are located on the same edge of the magnetic Brillouin zone, as shown in Fig.~\ref{fig:SIFig2}. For the FM edge, this yields two edges modes that overlap in a momenta region across the $\Gamma$-point [solid green line in Fig.~\ref{fig:SIFig2}(a)], while for the AFM edge, the two edges modes overlap in momentum regions to the left and right of the $\Gamma$-point [solid green lines in Fig.~\ref{fig:SIFig2}(b)].